\documentclass[journal]{IEEEtran}
\ifCLASSINFOpdf
\else
\fi
\usepackage{xcolor}
\usepackage[utf8]{inputenc}
\usepackage[english]{babel}
\usepackage{float}
\usepackage[pdftex]{graphicx}
\usepackage[caption=false]{subfig}

\usepackage{amsmath,amssymb,amsfonts}
\usepackage{amsmath}
\usepackage{textcomp}

\usepackage{amsthm}

\theoremstyle{definition}

\usepackage{algorithm}
\usepackage[noend]{algpseudocode}

\makeatletter
\def\BState{\State\hskip-\ALG@thistlm}
\makeatother

\algnewcommand\algorithmicswitch{\textbf{switch}}
\algnewcommand\algorithmiccase{\textbf{case}}
\algnewcommand\algorithmicassert{\texttt{assert}}
\algnewcommand\Assert[1]{\State \algorithmicassert(#1)}%
\algdef{SE}[SWITCH]{Switch}{EndSwitch}[1]{\algorithmicswitch\ #1\ \algorithmicdo}{\algorithmicend\ \algorithmicswitch}%
\algdef{SE}[CASE]{Case}{EndCase}[1]{\algorithmiccase\ #1}{\algorithmicend\ \algorithmiccase}%
\algtext*{EndSwitch}%
\algtext*{EndCase}%

 \usepackage{amssymb}
\usepackage{amsmath}

\usepackage{enumitem}
\usepackage{comment}
\begin{document}
%
\title{Optimising Energy Efficiency in UAV-Assisted Networks using Deep Reinforcement Learning}

%
%
%


\author{Babatunji Omoniwa,~Boris Galkin,~and~Ivana Dusparic
\thanks{All authors are with the CONNECT centre, School of Computer Science and Statistics, Trinity College Dublin, Ireland. e-mail: \{omoniwab,~galkinb,~duspari\}@tcd.ie. This work was supported, in part, by the Science Foundation Ireland (SFI) Grants No. 16/SP/3804 (Enable) and 13/RC/2077\_P2 (CONNECT Phase 2), as well as a research grant from SFI and the National Natural Science Foundation Of China (NSFC) under the SFI-NSFC Partnership Programme Grant Number 17/NSFC/5224.}}

%
%

\markboth{}%
{Omoniwa \MakeLowercase{\textit{et al.}}: Energy-efficiency optimisation in UAV-assisted networks}
%



\maketitle

\begin{abstract}

In this letter, we study the energy efficiency (EE) optimisation of unmanned aerial vehicles (UAVs) providing wireless coverage to static and mobile ground users. Recent multi-agent reinforcement learning approaches optimise the system's EE using a 2D trajectory design, neglecting interference from nearby UAV cells. We aim to maximise the system’s EE by jointly optimising each UAV's 3D trajectory, number of connected users, and the energy consumed, while accounting for interference. Thus, we propose a cooperative Multi-Agent Decentralised Double Deep Q-Network (MAD-DDQN) approach. Our approach outperforms existing baselines in terms of EE by as much as 55~--~80\%.

\end{abstract}


%
\IEEEpeerreviewmaketitle
\section{Introduction}
\label{sec:Introduction}
\IEEEPARstart{T}{he} deployment of unmanned aerial vehicles (UAVs) to provide wireless coverage to ground users has received significant research attention~\cite{Omoniwa_DQLCI_2021} -- \cite{Galkin2020UAVDeepLearning}. UAVs can play a vital role in supporting the Internet of Things (IoT) networks by providing connectivity to a large number of devices, static or mobile~\cite{Omoniwa_DQLCI_2021}. More importantly, UAVs have numerous real-world applications, ranging from assisted-communication in disaster-affected areas to surveillance, search and rescue operations~\cite{Zhang2021uav_emergency_comm},~\cite{ Xu2020_disaster_UAV}. Specifically, UAVs can be deployed in circumstances of network congestion or downtime of existing terrestrial infrastructure. Nevertheless, to provide ubiquitous services to dynamic ground users, UAVs require robust strategies to optimise their flight trajectory while providing coverage. As energy-constrained UAVs operate in the sky, they may be faced with the challenge of interference from nearby UAV cells or other access points sharing the same frequency band, thereby impacting the system's energy efficiency (EE)~\cite{Galkin2020UAVDeepLearning}.

There has been significant research effort on optimising EE in multi-UAV networks~\cite{Omoniwa_DQLCI_2021} --~\cite{Liu2018UAV}. The authors in~\cite{Mozaffari2017UAV} proposed an iterative algorithm to minimise the energy consumption of UAVs serving as aerial base stations to static ground users.
In~\cite{Ruan2018UAV}, a game-theoretic approach was proposed to maximise the system's EE while maximising the ground area covered by the UAVs irrespective of the presence of ground users. However, these works rely on a central ground controller for UAVs' decision making, thereby making it impractical to be deployed for emergencies due to the significant amount of exchanged information between the UAVs and the controller. Moreover, it may be difficult to track user locations in such a scenario. Machine learning is increasingly being used to address complex multi-UAV deployment problems. In particular, multi-agent reinforcement learning (MARL) approaches have been deployed in several works to optimise the system's EE. A distributed Q-learning approach~\cite{Omoniwa_DQLCI_2021} focused on optimising the energy utilisation of UAVs without considering the system's EE. To address this challenge, a deep reinforcement learning (DRL) approach~\cite{Galkin2020UAVDeepLearning} could be adopted. In our prior work~\cite{Galkin2022fixedWingUAV}, a DRL-based approach was proposed to optimise the EE of fixed-wings UAVs that move in circular orbits and are typically incapable of hovering like the rotary-winged UAVs. Moreover, the focus was on UAVs providing coverage to static ground users. The distributed DRL work in~\cite{Liu2020UAVdistributed} was an improvement on the centralised approach in~\cite{Liu2018UAV}, where all UAVs are controlled by a single autonomous agent. The authors in~\cite{Liu2020UAVdistributed},~\cite{Liu2018UAV} proposed a deep deterministic policy gradient (DDPG) approach to improve the system's EE as UAVs hover at fixed altitudes while providing coverage to static ground users in an interference-free network environment.
Although the approaches in~\cite{Liu2020UAVdistributed} and~\cite{Liu2018UAV} promise performance gains in terms of coverage score, they focus on the 2D trajectory optimisation of the UAVs serving static ground users.
\begin{figure}[!t]
\centering
\includegraphics[width=3.0in]{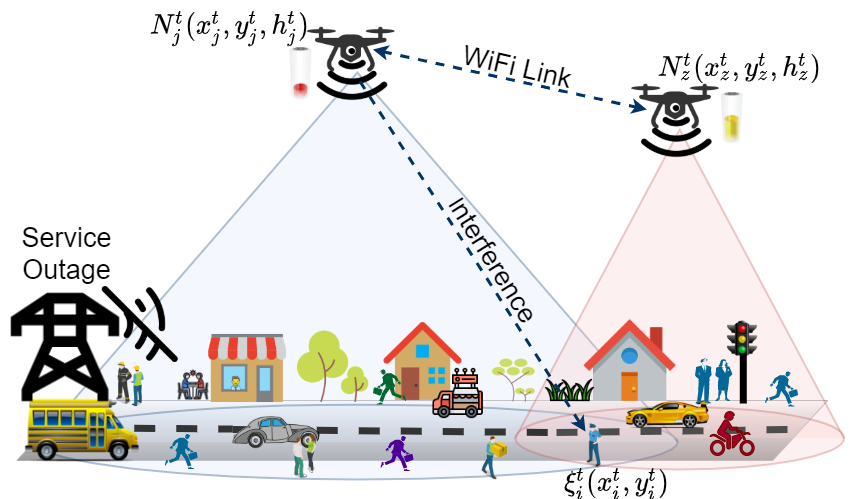}
 \caption{System model for UAVs serving static and mobile ground users.}
 \label{fig:systemmodel}
\end{figure}
Motivated by the research gaps above, we focus on maximising the system's EE by optimising the \textit{3D trajectory} of each UAV over a series of time steps, while taking into account the \textit{impact of interference} from nearby UAV cells and the coverage of both \textit{static and mobile ground users}. We propose a cooperative Multi-Agent Decentralised Double Deep Q-Network (MAD-DDQN) approach, where each agent's reward reflects the coverage performance in its neighbourhood. The MAD-DDQN approach maximises the system's EE without hampering performance gains in the network.
\section{System Model}
\label{section:SystemModel}
We consider a set of static and mobile ground users $\xi$ located in a given area, as shown in Figure \ref{fig:systemmodel}. Each user $i \in \xi$ at time $t$ is located in the coordinate $(x_i^t, y_i^t)$. We assume service unavailability from the existing terrestrial infrastructure due to disasters or increased network load. As such, a set $N$ of quadrotor UAVs are deployed within the area to provide wireless coverage to the ground users. A serving UAV $j \in N$ at time $t$ is located in the coordinate $(x_j^t, y_j^t, h_j^t)$. Without loss of generality, we assume a guaranteed line-of-sight (LOS) channel condition~\cite{Galkin2016UAV}, due to the aerial positions of the UAVs.
Signal-to-interference-plus-noise-ratio (SINR) is a measure of the signal quality. It can be defined as the ratio of the power of a certain signal of interest and the interference power from all the other interfering signals plus the noise power. Each user $i \in \xi$ in time $t$ can be connected to a single UAV $j \in N$ which provides the strongest downlink SINR. Thus, the SINR at time $t$ is expressed as~\cite{Omoniwa_DQLCI_2021},
\begin{equation}\label{eq:sinr}
  \gamma_{i,j}^t = \frac{\beta P (d_{i,j}^t)^{-\alpha}}{\Sigma_{z \in \chi_{int}} \beta P (d_{i,z}^t)^{-\alpha} + \sigma^2},
\end{equation}
where $\beta$ and $\alpha$ are the attenuation factor and path loss exponent that characterises the wireless channel, respectively. $\sigma^2$ is the power of the additive white Gaussian noise at the receiver, $d_{i,j}^t$ is the distance between the $i$ and $j$ at time $t$.~$\chi_{int} \in N$ is the set of interfering UAVs. $z$ is the index of an interfering UAV in the set~$\chi_{int}$. $P$ is the transmit power of the UAVs. We model the mobility of mobile users using the Gauss Markov Mobility (GMM) model~\cite{Camp2002}, which allows users to dynamically change their positions. UAVs must optimise their flight trajectory to provide ubiquitous connectivity to users. Given a channel bandwidth $B_w$, the receiving data rate of a ground user can be expressed using Shannon's equation~\cite{Galkin2020UAVDeepLearning},
\begin{equation}\label{eq:shannon}
  \mathbb{R}_{i,j}^t = B_w\log_2(1 + \gamma_{i,j}^t).
\end{equation}
In our interference-limited system, coverage is affected by the SINR. Hence, we compute the connectivity score of a UAV $j \in N$ at time $t$ as~\cite{Liu2020UAVdistributed},
\begin{equation}\label{eq:avgcoveragescore}
C_j^t = \sum_{\forall i \in \xi}w_j^t(i),
\end{equation}
where $w_j^t(i) \in [0, 1]$ denotes whether user $i$ is connected to UAV $j$ at time $t$. $w_j^t(i) = 1$ if $\gamma_{i}^t = \gamma_{i,j}^t > \gamma_{th}$, otherwise $w_j^t(i) = 0$, where $\gamma_{th}$ is the SINR predefined threshold. Likewise $\mathbb{R}_{i,j}^t = 0$ if user $i$ is not connected to UAV $j$.

During flight operations, a UAV $j \in N$ at time $t$ expends energy $e_j^t$. A UAVs' total energy $e_T$ is expressed as the sum in propulsion~$e_P$ and communication~$e_C$ energies, $e_T = e_P + e_C$. Since $e_C$ is practically much smaller than $e_P$, i.e., $e_C \ll e_P$ \cite{Omoniwa_DQLCI_2021}, we ignore $e_C$. A closed-form analytical propulsion power consumption model for a rotary-wing UAV at time $t$ is given as \cite{Zeng2019UAV},
\vspace{-2mm}
\begin{equation}\label{eq:powertime}
P(t) = \kappa_0 \Big(1 + \frac{3V^2}{U^2_{tip}}\Big)  + \kappa_i \Big( \sqrt{1 + \frac{V^4}{4v_0^4}} + \frac{V^2}{2v_0^2}\Big)^{\frac{1}{2}} + \frac{\rho}{2} \nu s A V^3,\\
\end{equation}
where $\kappa_0$ and $\kappa_i$ are the UAVs' flight constants (e.g., rotor radius or weight), $U_{tip}$ is the rotor blade's tip speed, $v_0$ is the mean hovering velocity, $\nu$ is the drag ratio, $s$ is the rotor solidity, $A$ is the rotor disc area, $V$ is the UAVs' speed at time $t$ and $\rho$ is the air density. In particular, we take into account the basic operations of the UAV, such as, hovering and acceleration. Therefore, we can derive the average propulsion power over all time steps as $\frac{1}{T}\sum_{t=1}^{T}P(t)$, and the total consumed energy of a UAV $j$ is given as~\cite{Omoniwa_DQLCI_2021},
\vspace{-1mm}
\begin{equation}\label{eq:propenergy}
e_{j}^t = \delta_t \sum_{t=1}^{T}P(t),
\vspace{-1mm}
\end{equation}
where $\delta_t$ is the duration of each time step. The EE at time $t$ can be expressed as the ratio of the total data throughput and the total energy consumed by all UAVs, expressed as,
\begin{equation}\label{eq:energyEfficiency}
    \eta_{t} = \frac{\sum\limits_{j \in N}\sum\limits_{i \in \xi} \mathbb{R}_{i,j}^t}{\sum\limits_{j \in N}e_j^t}.
\end{equation}
\section{Multi-Agent Reinforcement Learning Approach for Energy Efficiency Optimisation}
\label{section:mad-dqn}
In this section, we formulate the problem and propose a our MAD-DDQN algorithm to improve the trajectory of each UAV in a manner that maximises the total system's EE.
\vspace{-1mm}
\begin{figure*}[!htbp]
\centering
\includegraphics[width=6.4in]{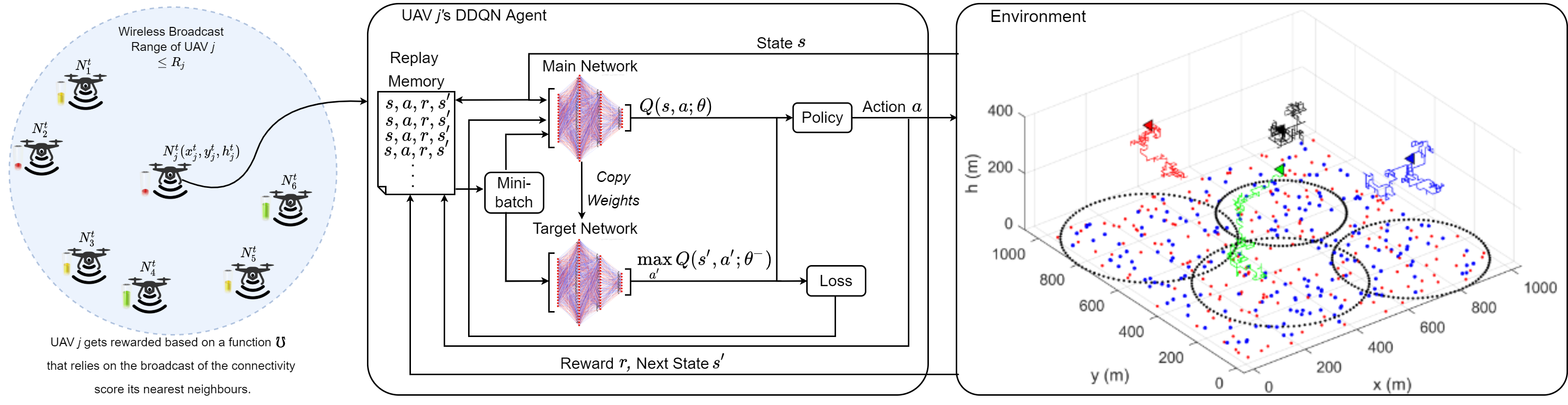}
\caption{Multi-agent decentralised double deep Q-network framework where each UAV $j$ equipped with a DDQN agent interacts with its environment. The environment shows the simulation snapshot of UAVs providing wireless coverage to 200 static (blue) and 200 mobile (red) ground users with flight trajectories. On the left shows the broadcast range of UAV $j$ in a multi-UAV scenario, where UAVs broadcast their telemetry information to nearest neighbours}
\label{fig:madddqn}
\end{figure*}
\subsection{Problem Formulation}
Our objective is to maximise the total system's EE by jointly optimising its 3D trajectory, number of connected users, and the energy consumed by the UAVs serving ground users under a strict energy budget. Maximising the number of connected users $C_j^t$ will maximise the total amount of data $\sum\limits_{i \in \xi} \mathbb{R}_{i,j}^t$ the UAV $j$ will deliver in time step $t$ which, for a given amount of consumed energy $e_j^t$, will also maximise the EE $\eta_{t}$. Therefore, the optimisation problem can be formulated as,
\begin{subequations}
\begin{align} \label{eq:problem_statement}
&\underset{\forall j \in N:~\mathbf{x_j^t,~y_j^t,~h_j^t,~e_j^t,~C_j^t}}{\max}~\eta_{tot} =~\sum\limits_{t=1}^T \eta_{t} \\
\text{s.t.} \quad &  \gamma_{i,j}^t \geq \gamma_{th},\quad \forall w_j^t(i) \in [0, 1],~i,~j,~t, \\
&e_j^t \leq e_{\max},\quad \quad \quad \quad \quad \quad \quad \forall j,~t,\\
&x_{\min}~\leq~x_j^t~\leq x_{\max},\quad \quad ~~\forall j,~t,\\
&y_{\min}~\leq~y_j^t~\leq y_{\max},\quad \quad ~~~\forall j,~t,\\
&h_{\min} \leq h_j^t \leq h_{\max},\quad \quad \quad ~~\forall j,~t,
\end{align}
\end{subequations}
where $e_{\max}$ is the maximum UAV energy level, $x_{min}$, $y_{min}$, $h_{min}$ and $x_{max}$, $y_{max}$, $h_{max}$ are the minimum and maximum 3D coordinates of $x$, $y$ and $h$, respectively.
As multiple wireless transmitters sharing the same frequency band are in close proximity to one another the possibility of interference is significantly increased.
The computational complexity of problem (\ref{eq:problem_statement}) is known to be NP-complete~\cite{Liu2019UAVqlearning}. The problem (\ref{eq:problem_statement}) is non-convex, thus having multiple local optimum. For this reason, solving (\ref{eq:problem_statement}) with conventional optimization approaches is challenging~\cite{Omoniwa_DQLCI_2021},~\cite{Liu2019UAVqlearning}.
Specifically, the problem (\ref{eq:problem_statement}) will become more complex as more UAVs are deployed in a shared wireless environment, hence it is challenging to find the optimal cooperative strategies to improve the system's EE while completing the coverage tasks under dynamic settings. This is often because UAVs may become selfish and pursue the goal of improving their individual EE while minimising the communication outage and energy consumption, rather than the collective goal of maximising the system's EE. In such cases, cooperative MARL approaches may be suitable when individual and collective interests of UAVs conflict. Deep RL has been shown to perform well in decision-making tasks in such a dynamic environment~\cite{Zhang2021uavBS}. Hence, we adopt a cooperative deep MARL approach to solve the system's EE optimisation problem.
\begin{algorithm}
\footnotesize
\caption{Double Deep Q-Network (DDQN) for Agent $j$}\label{Algorithm_DDQN}
\begin{algorithmic}[1]
\State Input: UAV3Dposition, ConnectivityScore, InstantaneousEnergyConsumed $\in S$ and Output: Q-values corresponding to each possible action~$(+x_s,~0,~0)$, $(-x_s,~0,~0)$, $(0,+y_s,~0)$, $(0,-y_s,~0)$, $(0,~0,+z_s)$, $(0,~0,-z_s)$, $(0,~0,~0)$~$\in A_j$
\State \parbox[t]{0.8\linewidth}{$\mathcal{D}$ -- empty replay memory,  $\theta$ -- initial network parameters, $\theta^{-}$ -- copy of $\theta$, $\mathcal{N}_r$ -- maximum size of replay memory, \\$\mathcal{N}_b$ -- batch size, $\mathcal{N}^{-}$ -- target replacement frequency.}

\State $s$ $\gets$ initial state, maxStep $\gets$ maximum number of steps in the episode
\While{\emph{goal} not Reached and Agent \emph{alive} and maxStep not reached}
\State \emph{s} $\leftarrow$ MapLocalObservationToState(\emph{Env})
\State \Comment{\parbox[t]{0.8\linewidth}{\texttt{Execute $\epsilon$-greedy method based on $\pi_j$}}}
\State \emph{a} $\leftarrow$ DeepQnetwork.SelectAction(\emph{s})
\State \Comment{\parbox[t]{0.8\linewidth}{\texttt{Agent executes action in state $s$}}}
\State  \emph{a}.execute(\emph{Env})
\If { \emph{a}.execute(\emph{Env}) is True}
\State \Comment{\parbox[t]{0.8\linewidth}{\texttt{Map sensed observations to new state $s'$}}}
\State Env.UAV3Dposition \cite{Liu2019UAVqlearning}
\State Env.ConnectivityScore  (\ref{eq:avgcoveragescore})
\State Env.InstantaneousEnergyConsumed (\ref{eq:propenergy})
\EndIf
\State \emph{r} $\leftarrow$ Env.RewardWithCooperativeNeighbourFactor (\ref{eqnreward})
\State \Comment{\parbox[t]{0.8\linewidth}{\texttt{ Execute UpdateDDQNprocedure()}}}
\State Sample a minibatch of $\mathcal{N}_b$ tuples $(s, a, r, s') \sim Unif(\mathcal{D})$ \label{Algline:dqn_start}
\State Construct target values, one for each of the $\mathcal{N}_b$ tuples:
\State Define $a^{max} (s'; \theta) =  \arg \max_{a'} Q(s', a'; \theta)$
\If {\emph{$s'$} is Terminal}
\State $y_j =  r$
\Else
\State $y_j =  r + \gamma Q(s', a^{max} ((s'; \theta); \theta^{-}) $
\EndIf
\State Apply a gradient descent step with loss $\parallel y_j - Q(s, a; \theta)\parallel^2$
\State Replace target parameters $\theta^{-} \gets \theta$ every $\mathcal{N}^{-}$ step \label{Algline:dqn_stop}
\EndWhile
\State \textbf{endwhile}
\end{algorithmic}
\end{algorithm}
\vspace{-2mm}
\subsection{Cooperative Multi-Agent Decentralised Double Deep Q-Network (MAD-DDQN)}
We propose a cooperative MAD-DDQN approach, where each agent's reward reflects the coverage performance in its neighbourhood. Here, each UAV is controlled by a Double Deep Q-Network (DDQN) agent that aims to maximise the system's EE by jointly optimising its 3D trajectory, number of connected users, and the energy consumed. We assume the agents interact with each other in a shared and dynamic environment, which may lead to learning instabilities due to conflicting policies from other agents. From Algorithm \ref{Algorithm_DDQN}, Agent $j$ follows an $\epsilon$--greedy policy by executing an action $a$, transiting from state $s$ to a new state $s'$ and receiving a reward reflecting the coverage performance in its neighbourhood in (\ref{eqnreward}), after which DDQN procedure described on line \ref{Algline:dqn_start}--\ref{Algline:dqn_stop} optimises the agent's decisions. We explicitly define the states, actions, and reward as follows: \begin{itemize}[leftmargin=*]
    \item \textbf{State space}: We consider the three-dimensional (3D) position of each UAV~\cite{Liu2019UAVqlearning}, the connectivity score and the UAV's instantaneous energy level at time $t$, expressed as a tuple,~\textlangle{}$x^t : \{0, 1, ..., x_{max}\},~y^t : \{0, 1, ..., y_{max}\},~h^t : \{h_{min}, ..., h_{max}\},~C_t, ~e_t $\textrangle{}.
    \item \textbf{Action space}: At each time-step~$t \in T$, each UAV takes an action by changing its direction along the 3D coordinates. Unlike our closest related work and the evaluation baseline~\cite{Liu2020UAVdistributed}, we discretise the agent's actions following the design from ~\cite{Omoniwa_DQLCI_2021} and~\cite{Liu2019UAVqlearning}, as follows:~$(+x_s,~0,~0)$, $(-x_s,~0,~0)$, $(0,+y_s,~0)$, $(0,-y_s,~0)$, $(0,~0,+z_s)$, $(0,~0,-z_s)$ and $(0,~0,~0)$. Our rationale to discretise the action space was to ensure quick adaptability and convergence of the agents.
    \item \textbf{Reward}: The agent's goal is to learn a policy that implicitly maximises the system's EE by jointly minimising the ground users outage and total UAVs energy consumption. Hence, we introduce a shared cooperative factor~$\mho$ to shape the reward formulation of each agent $j$ in each time-step $t \in T$ given as,
    \begin{equation}\label{eqnreward}
        \mathcal{R}_j^t =
        \begin{cases}
          ~~\mho + \omega + 1, & \text{if}\ C_j^t > ~C_j^{t - 1}\\
          ~~\mho + \omega, & \text{if}\ C_j^{t} = ~C_j^{t - 1}\\
          ~~\mho + \omega - 1, & \text{otherwise,}
        \end{cases}
        \vspace{-1mm}
    \end{equation}
    where $C_j^{t}$ and $C_j^{t - 1}$ are the connectivity score in present and previous time-step, respectively. $\omega~=~\frac{e_j^{t - 1} - e_j^{t}}{e_j^{t} + ~e_j^{t - 1}}$, where $e_j^{t}$ and $e_j^{t - 1}$ are the instantaneous energy consumed by agent $j$ in present and previous time-step, respectively. To enhance cooperation, we assign each agent a `$+1$' incentive from its neighbourhood via a function~$\mho$ only when the overall connectivity score, which is the total number of connected users by UAVs in its locality in the present time-step $C_{t}^o$ exceeds that in the previous time-step $~C_{t - 1}^o$, otherwise the agent receives a `$-1$' incentive. We compute~$\mho$ as,
    \begin{equation}\label{coop_factor}
    \mho =
    \begin{cases}
      ~+1, & \text{if}\ C_{t}^o~> ~C_{t - 1}^o\\
      ~-1, & \text{otherwise.}
    \end{cases}
    \end{equation}
\end{itemize}
\begin{figure*}[!htbp]
\begin{minipage}[b]{0.33\linewidth}
\centering
\subfloat[Subfigure 1 list of figures text][Energy efficiency $\eta$ vs. number of UAVs.]{\includegraphics[width=\textwidth]{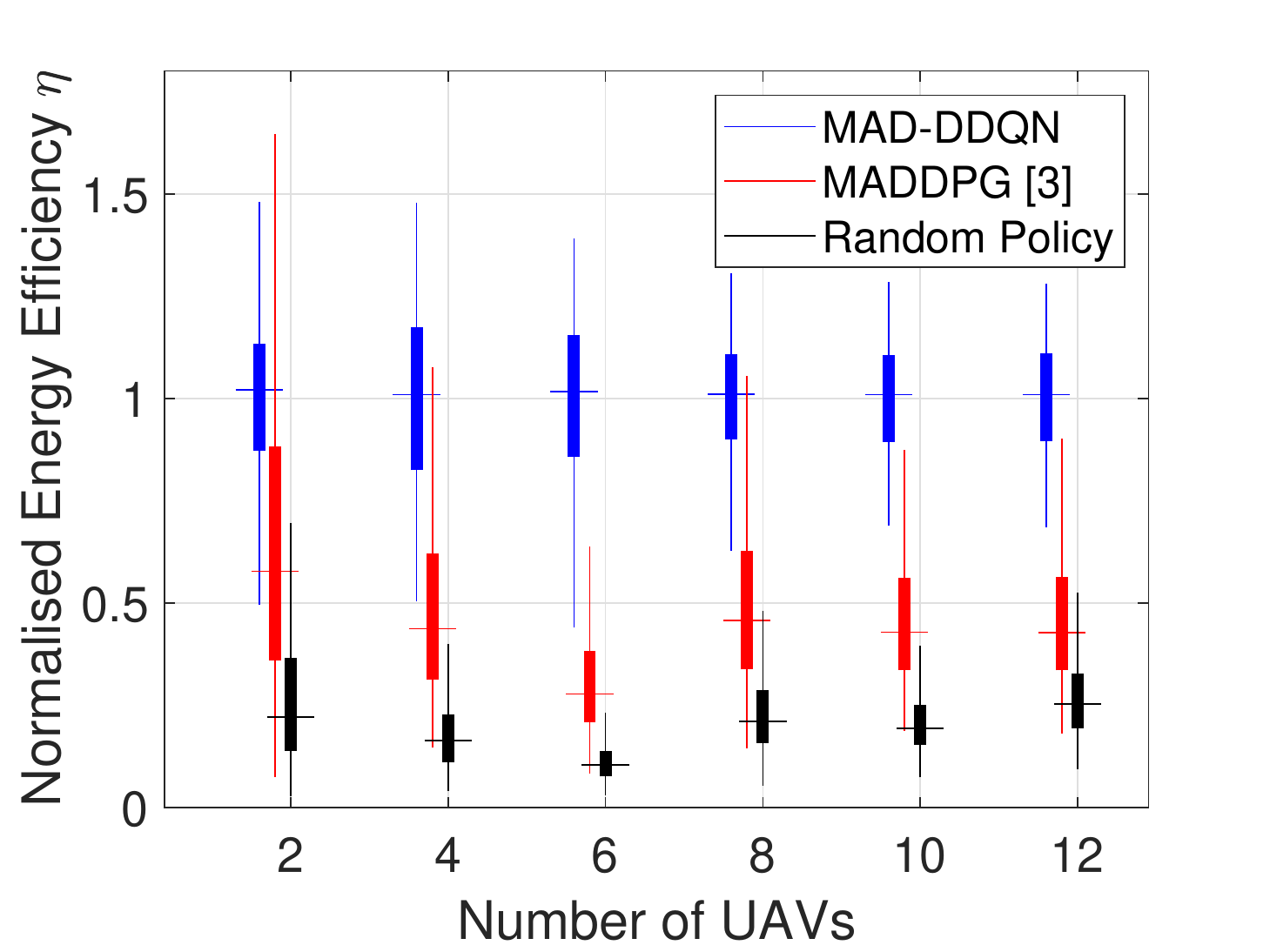}
\label{subfig:EEvsNumUAV_letters}}
\end{minipage}
\begin{minipage}[b]{0.33\linewidth}
\centering
\subfloat[Subfigure 2 list of figures text][Ground users outage vs. number of UAVs.]{
\includegraphics[width=\textwidth]{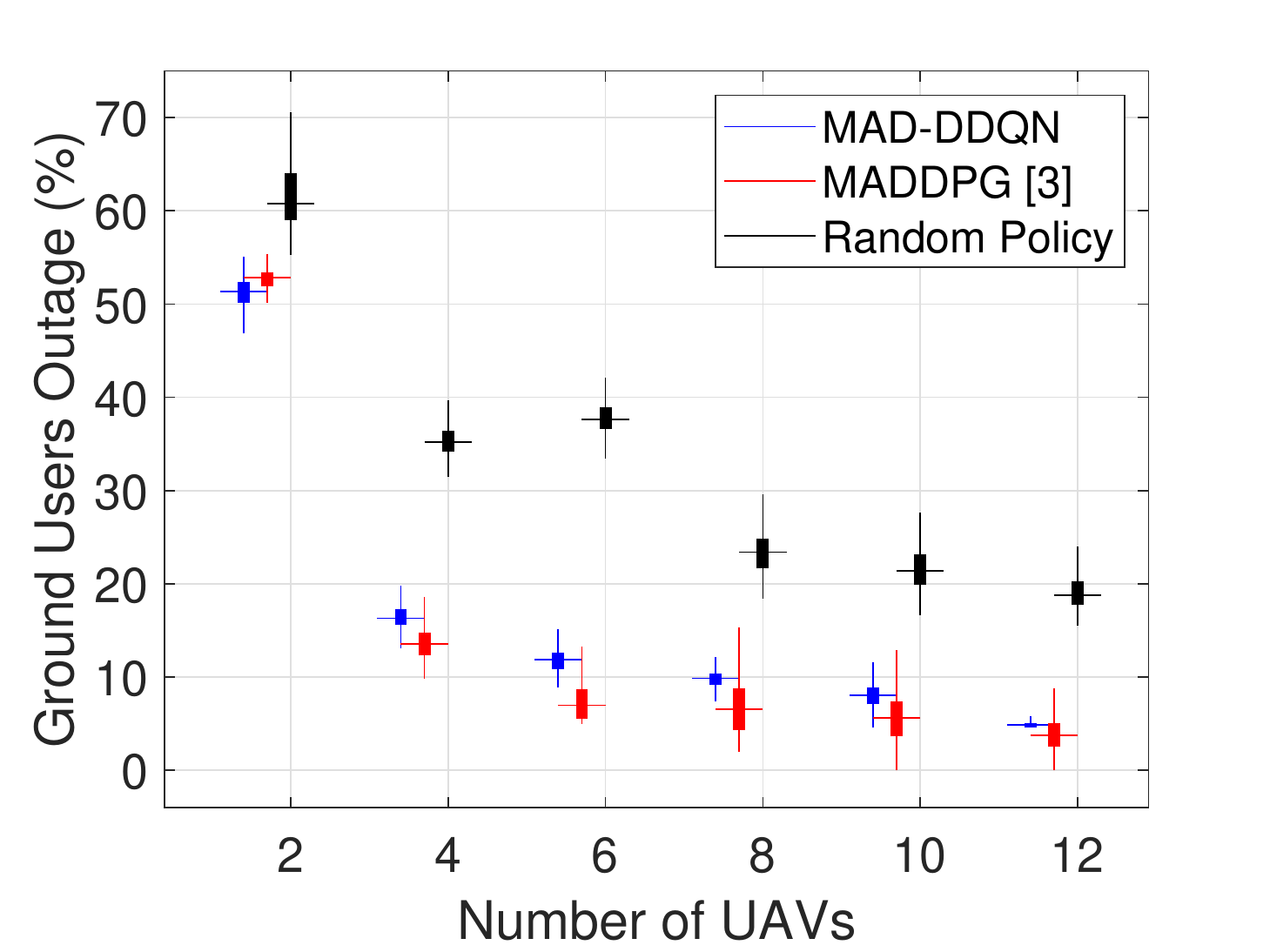}
\label{subfig:outagevsNumUAV}}
\end{minipage}
\begin{minipage}[b]{0.33\linewidth}
\centering
\subfloat[Subfigure 3 list of figures text][Total energy consumed vs. number of UAVs.]{
\includegraphics[width=\textwidth]{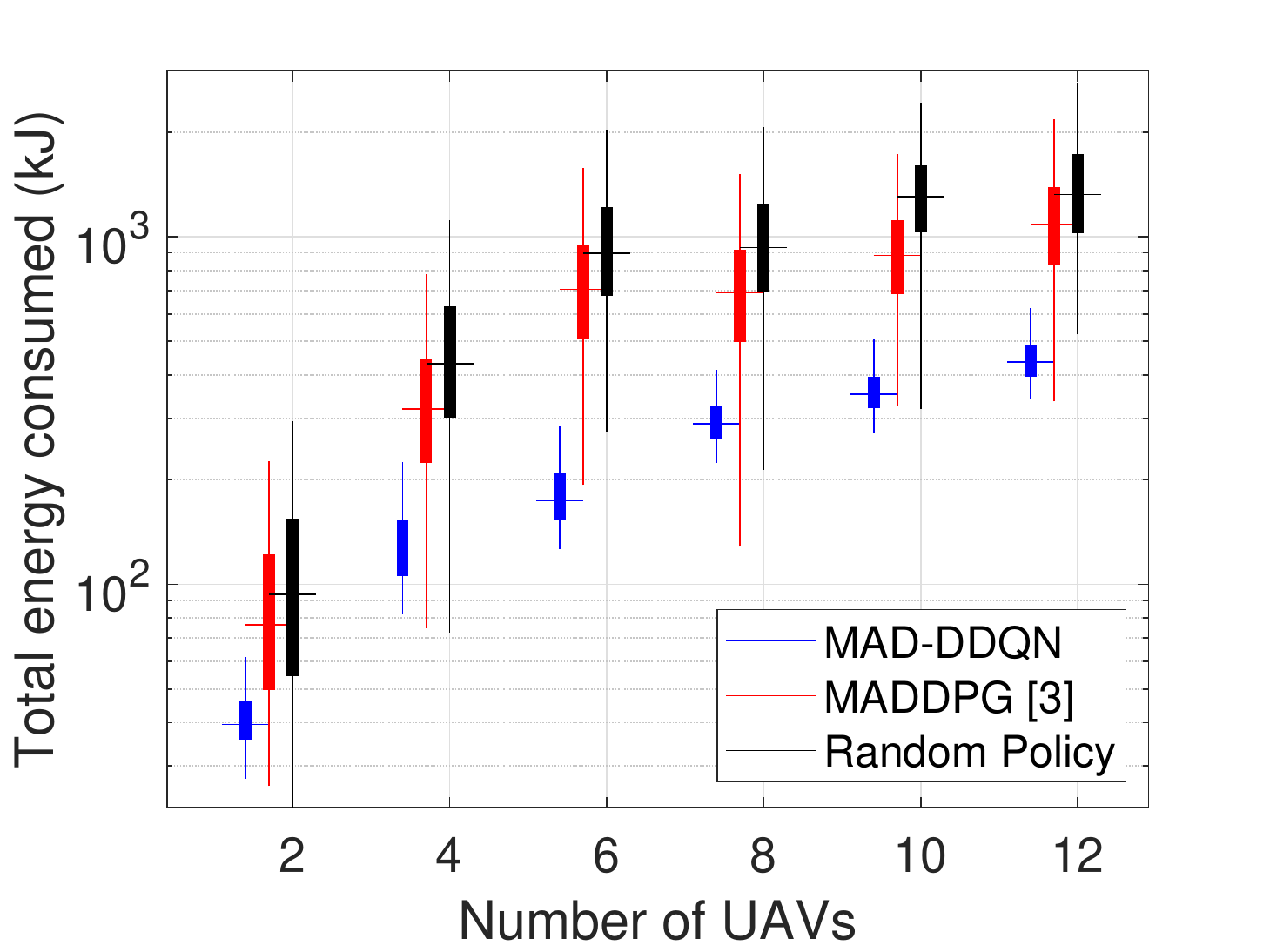}
\label{subfig:energyvsNumUAV_letters}}
\end{minipage}
\caption{Impact of number of deployed UAVs on the UAVs' EE, ground users outage and total energy consumption under dynamic network conditions with 400 ground users deployed in a 1 km$^2$ area, with results from 2000 runs of MC trials.}
\label{fig: comp_metrics_number_letters}
\vspace{-3mm}
\end{figure*}
\vspace{-7mm}
\subsection{DDQN Implementation}
The neural network (NN) architecture of Agent $j$'s DDQN shown in Figure \ref{fig:madddqn} comprises of a 5-dimensional state space input vector, densely connected to 2 layers with 128 and 64 nodes, with each using a rectified linear unit (ReLU) activation function, leading to an output layer with 7 dimensions. Our decentralised approach assume agents to be independent learners. Following the analysis presented in \cite{Hribar2021_complexity_DRL}, the computational complexity of the NN architecture used in the MAD-DDQN is approximately $\mathcal{O}(D_s K W)$ with an average response time of 5.6 ms, while that of our closest related work and the evaluation baseline~\cite{Liu2020UAVdistributed} (MADDPG) is approximately $\mathcal{O}(D_s K W) + \mathcal{O}((D_a + D_s) K W)$ with an average response time of 7.4 ms, where $D_s$ is the dimension of the state space, $D_a$ is the dimension of the action space, $K$ is the number of layers, $W$ is the number nodes in each hidden layer. \newline \indent In the training phase, given the state information as input, Agent $j$ trains the main network to make better decision by yielding Q-values corresponding to each possible action as output. The maximum Q-value obtained determines the action the agent executes. At each time step Agent $j$ observes its present state $s$ and updates it’s trajectory by selecting an action $a$ in accordance with its policy. Following its action in time step $t$, Agent $j$ observes a reward $r$ which is defined in (\ref{eqnreward}), and transits to a new state $s'$. The information $(s,a,r,s')$ is inputted in the replay memory as shown in Figure \ref{fig:madddqn}. Agent $j$ then samples the random mini-batch from the replay memory and uses the mini-batch to obtain $y_j$. The optimisation is performed with $L(\theta)$ and $\theta$ updated accordingly. In every 100th time step, the target Q-network updates the parameters $\theta^-$ with the same parameters $\theta$ of the main network. For the training, the memory size was set to 10,000, and the mini-batch size was set to 1024. The optimisation is performed using a variant of the stochastic gradient descent called RMSprop to minimise the loss following the methodology described in \cite[Chapter 4]{Francois-Lavet2018}. The learning rate and discount factor were set to 0.0001 and 0.95, respectively. We train the Q-networks by running multiple episodes, and at each training step the $\epsilon$-greedy policy is used to have a balance between exploration and exploitation~\cite{Francois-Lavet2018}. In the $\epsilon$-greedy policy, the action is randomly selected with $\epsilon$ probability, whereas the action with the largest action value is selected with a probability of $1-\epsilon$. The initial value of $\epsilon$ was set to 1 and linearly decreased to 0.01.

\begin{table}[!t]
\small
\centering
\caption{Simulation Parameters}
\label{table:parameters}
\begin{tabular}{|l|l|}
  \hline
 \textit{Parameters} & \textit{Value} \\
  \hline \hline
  Software platform/Library & Python 3.7.4/PyTorch 1.8.1\\
       Optimiser/Loss function & RMSprop/MSELoss\\
       Learning rate/Discount factor & 0.0001/0.95\\
       Hidden layers/Activation function & 2 (128, 64)/ReLu\\
       Replay memory size/Batch size & 10,000/1024\\  
       Policy/Episodes/maxStep & $\epsilon$-greedy/250/1500\\
       No. of ground users/Model & 400/GMM\\
       Ground user direction/Velocity & [0, 2$\pi$]/[0, 15]~mps\\
       Number of UAVs/Weight per UAV & [2--12]/16 kg\\
       Nominal battery capacity & 16,000 mAh\\
       Maximum transmit power~\cite{Liu2019UAVqlearning} & 20 dBm\\
       Noise power/SINR threshold~\cite{Mozaffari2017UAV} & -130 dBm/5 dB\\
       Bandwidth~\cite{Liu2019UAVqlearning} & 1 MHz\\
       Pathloss exponent~\cite{Mozaffari2017UAV},~\cite{Liu2019UAVqlearning} & 2\\
       UAV step distance ($\forall~x_s, y_s, z_s$) & [0--20] m\\
      \hline
 \end{tabular}
 \end{table}
\section{Evaluation and Results}
\label{section:SimresultSection}
In this section, we verify the effectiveness of the proposed MAD-DDQN approach against the following baselines: (\textit{i}) the random policy; and (\textit{ii}) the MADDPG~\cite{Liu2020UAVdistributed} approach that considers a 2D trajectory optimisation while neglecting interference from nearby UAV cells. Simulation parameters are presented in Table~\ref{table:parameters}. We simulate a varying number of UAVs ranging from 2 to 12 to serve both static and mobile ground users in a 1000$\times$1000 $m^2$ area as shown in Figure \ref{fig:madddqn}. We perform 2000 runs of Monte-Carlo (MC) trials over trained episodes. In Figure~\ref{fig: comp_metrics_number_letters}, we compare the MAD-DDQN approach with baselines to evaluate the impact of different number of deployed UAVs on the EE, ground users outage and total energy consumption. Due to baseline MADDPG approach taking significantly longer to converge (learn suitable behaviours), to achieve a fair comparison,  Figure~\ref{fig: comp_metrics_number_letters} compares the performance after training the MAD-DDQN approach for 250 episodes and the MADDPG approach for 2000 episodes.


Since we focus on comparing the EE values rather than showing their absolute values, we normalise the EE values with respect to the mean values of the proposed MAD-DDQN approach. From Figure \ref{subfig:EEvsNumUAV_letters}, we observe that the MAD-DDQN approach consistently outperforms the random policy and MADDPG approaches across the entire range of UAVs deployment by approximately 80\% and~55\%, respectively.
\begin{figure}[!t]
\centering
\includegraphics[width=3.0in]{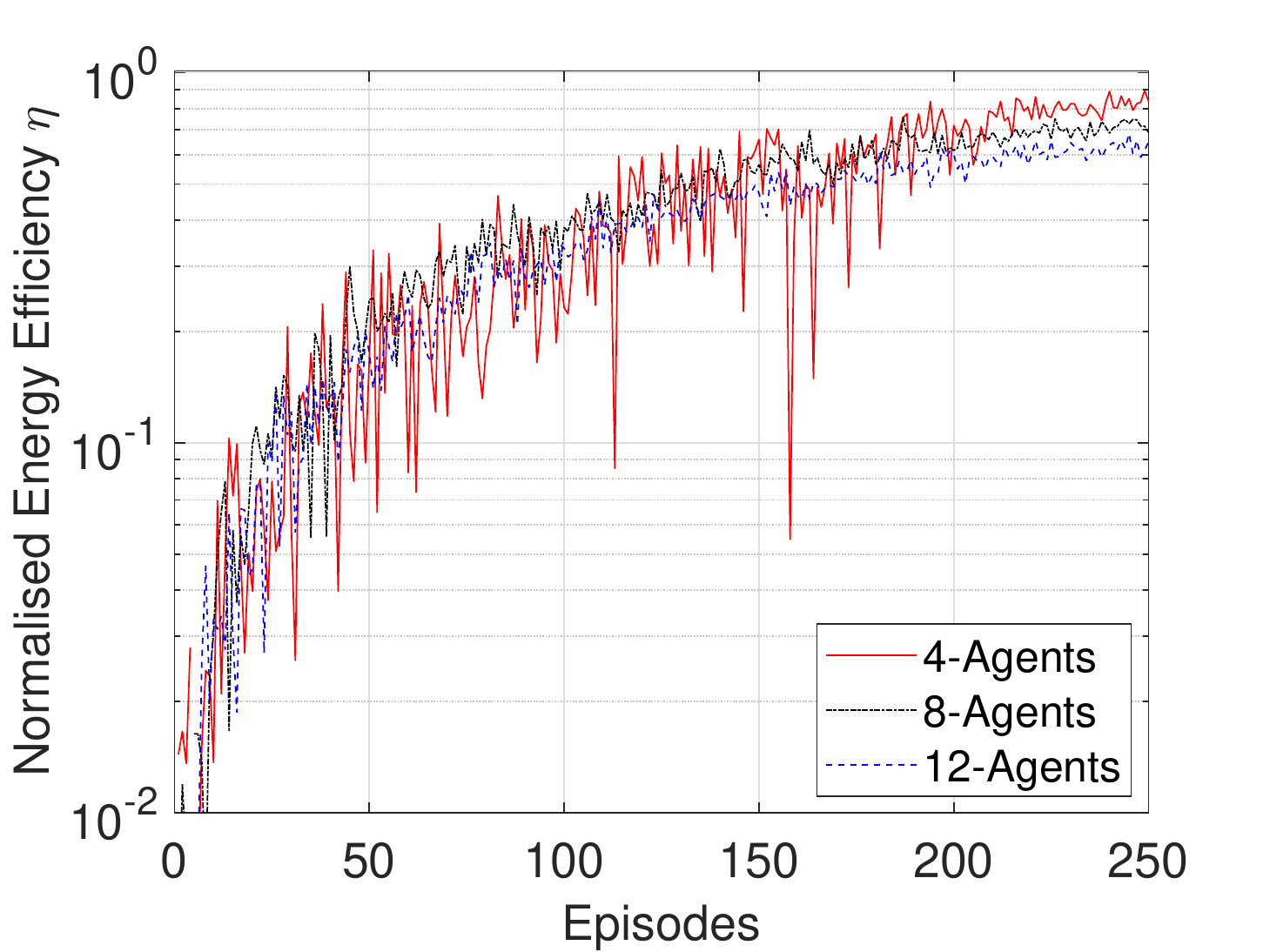}
\caption{Energy efficiency $\eta$ vs. learning episodes showing the convergence of MAD-DDQN while varying the number of agents.}
\label{Fig:EEvsEpiLetters}
\end{figure}
Interestingly, we see a marginally better performance by the MADDPG approach over the MAD-DDQN approach in  minimising the outages experienced by ground users by about 2\%, as shown in Figure \ref{subfig:outagevsNumUAV}. However, the slight performance gain by the MADDPG comes at a huge computational training cost which is 8 times higher than the MAD-DDQN approach. Intuitively, the MAD-DDQN approach hides redundant information about the environment through discretisation of the agent’s action space, which makes the MAD-DDQN approach require less experience to successfully learn a policy than the MADDPG approach. On the other hand, the random policy performed worst among the approaches in reducing connection outages, emphasizing the relevance of strategic decision making in MARL problems. Figure \ref{subfig:energyvsNumUAV_letters} clearly shows that the proposed approach significantly minimises the total energy consumed by all UAVs as compared to the baselines. Although the MADDPG approach performs slightly better at reducing outages than our approach, our MAD-DDQN approach is significantly more energy efficient, hereby implying the MADDPG approach trades energy consumption for improved coverage of ground users. In Figure \ref{Fig:EEvsEpiLetters}, we show the plot of the EE versus the learning episodes while varying the number of agents to demonstrate the convergence behaviour of the MAD-DDQN approach. We observe a steady decrease in the converged values of the EE while increasing the number of UAVs because the system becomes more unstable with more UAVs, thereby decreasing the system throughput as interference increases. Overall, the cooperative MAD-DDQN approach shows convergence in the system's EE irrespective of the number of UAVs deployed in the network.
\section{Conclusion}
\label{conclusion}
In this letter, we propose a MAD-DDQN approach to optimise the EE of a fleet of UAVs serving static and mobile ground users in an interference-limited environment. The MAD-DDQN approach guarantees quick adaptability and convergence, thereby allowing agents to learn policies that maximise the total system's EE by jointly optimising its 3D trajectory, number of connected users, and the energy consumed by the UAVs serving ground users under a strict energy budget. Extensive simulation
results have demonstrated that the MAD-DDQN approach significantly outperforms the random policy and a state-of-the-art decentralised MARL solution in terms of EE without degrading coverage performance in the network.


\ifCLASSOPTIONcaptionsoff
  \newpage
\fi


\begin{thebibliography}{1}
\bibitem{Omoniwa_DQLCI_2021}
B. Omoniwa, B. Galkin and I. Dusparic, ``Energy-aware optimization of UAV base stations placement via decentralized multi-agent Q-learning,'' \emph{2022 IEEE 19th Annual Consumer Communications \& Networking Conference (CCNC),} Jan. 2022, pp. 216-222.

\bibitem{Mozaffari2017UAV}
M. Mozaffari, W. Saad, M. Bennis and M. Debbah, ``Mobile Unmanned Aerial Vehicles (UAVs) for Energy-Efficient Internet of Things Communications,'' \emph{IEEE Transactions on Wireless Communications,} vol. 16, no. 11, pp. 7574-7589, Nov. 2017.

\bibitem{Liu2020UAVdistributed}
C. H. Liu, X. Ma, X. Gao and J. Tang, ``Distributed Energy-Efficient Multi-UAV Navigation for Long-Term Communication Coverage by Deep Reinforcement Learning,'' \emph{IEEE Transactions on Mobile Computing,} vol. 19, no. 6, pp. 1274-1285, June 2020.

\bibitem{Ruan2018UAV}
L. Ruan et al., ``Energy-efficient multi-UAV coverage deployment in UAV networks: A game-theoretic framework,'' \emph{China Communications,} vol. 15, no. 10, pp. 194-209, Oct. 2018.

\bibitem{Liu2018UAV}
C. H. Liu, Z. Chen, J. Tang, J. Xu and C. Piao, ``Energy-Efficient UAV Control for Effective and Fair Communication Coverage: A Deep Reinforcement Learning Approach,'' \emph{IEEE Journal on Selected Areas in Communications,} vol. 36, no. 9, pp. 2059-2070, Sept. 2018.

\bibitem{Liu2019UAVqlearning}
X. Liu, Y. Liu and Y. Chen, ``Reinforcement Learning in Multiple-UAV Networks: Deployment and Movement Design,'' \emph{IEEE Transactions on Vehicular Technology,} vol. 68, no. 8, pp. 8036-8049, Aug. 2019.

\bibitem{Galkin2020UAVDeepLearning}
B. Galkin, E. Fonseca, R. Amer, L. A. DaSilva and I. Dusparic, ``REQIBA: Regression and Deep Q-Learning for Intelligent UAV Cellular User to Base Station Association,'' \emph{IEEE Transactions on Vehicular Technology,} vol. 71, no. 1, pp. 5-20, Jan. 2022.

\bibitem{Zhang2021uav_emergency_comm}
C. Zhang, M. Dong and K. Ota, ``Heterogeneous Mobile Networking for Lightweight UAV Assisted Emergency Communication,'' \emph{IEEE Transactions on Green Communications and Networking,} vol. 5, no. 3, pp. 1345-1356, Sept. 2021.

\bibitem{Xu2020_disaster_UAV}
J. Xu, K. Ota and M. Dong, ``Big Data on the Fly: UAV-Mounted Mobile Edge Computing for Disaster Management,'' \emph{IEEE Transactions on Network Science and Engineering,} vol. 7, no. 4, pp. 2620-2630, Oct.-Dec. 2020.

\bibitem{Galkin2022fixedWingUAV}
B. Galkin, B. Omoniwa, and I. Dusparic, ``Multi-Agent Deep Reinforcement Learning For Optimising Energy Efficiency of Fixed-Wing UAV Cellular Access Points,'' \emph{ICC 2022 - IEEE International Conference on Communications, (to appear), arXiv:2111.02258,} May 2022.

\bibitem{Galkin2016UAV}
B. Galkin, J. Kibilda and L. A. DaSilva, ``Deployment of UAV-mounted access points according to spatial user locations in two-tier cellular networks,'' \emph{2016 Wireless Days (WD),} 2016, pp. 1-6.

\bibitem{Camp2002}
T. Camp, J. Boleng, V. Davies, ``A Survey of Mobility Models for Ad Hoc Network Research,'' \emph{Wireless Communication \& Mobile Computing (WCMC): Special issue on Mobile Ad Hoc Networking: Research, Trends and Applications,} 2002, pp. 483-502.

\bibitem{Zeng2019UAV}
Y. Zeng, J. Xu and R. Zhang, ``Energy Minimization for Wireless Communication With Rotary-Wing UAV,'' \emph{IEEE Transactions on Wireless Communications,} vol. 18, no. 4, pp. 2329-2345, April 2019.

\bibitem{Zhang2021uavBS}
M. Zhang, S. Fu and Q. Fan, ``Joint 3D Deployment and Power Allocation for UAV-BS: A Deep Reinforcement Learning Approach,'' \emph{IEEE Wireless Commun. Lett.,} vol. 10, no. 10, pp. 2309-2312, Oct. 2021.

\bibitem{Hribar2021_complexity_DRL}
J. Hribar, A. Marinescu, A. Chiumento and L. A. DaSilva, ``Energy Aware Deep Reinforcement Learning Scheduling for Sensors Correlated in Time and Space,'' \emph{IEEE Internet of Things Journal,} doi: 10.1109/JIOT.2021.3114102.

\bibitem{Francois-Lavet2018}
V. Franc\c{o}is-Lavet, P. Henderson, R. Islam, M. G. Bellemare, and J. Pineau, ``An Introduction to Deep Reinforcement Learning,'' \emph{Foundations and Trends in Machine Learning,} vol. 11, no. 3-4, 2018.



\end{thebibliography}
\end{document}